\documentclass[twocolumn,superscriptaddress,showpacs,preprintnumbers,floatfix,aps]{revtex4-1}

\usepackage[pdftex]{color,graphicx}
\usepackage{latexsym,amsmath,amssymb,gensymb}
\usepackage{hyperref}
\usepackage[flushleft]{threeparttable}

\begin{document}

\title{Molecular insights on Poly(N-isoproylacrylamide) coil-to-globule transition \\induced by pressure}
 
\author{Letizia Tavagnacco}
\affiliation{CNR-ISC and Department of Physics, Sapienza University of Rome,\\Piazzale A. Moro 2, 00185, Rome, Italy.}

\author{Ester Chiessi}
\email[Corresponding author: ]{ester.chiessi@uniroma2.it}
\affiliation{Department of Chemical Sciences and Technologies, University of Rome Tor Vergata,\\Via della Ricerca Scientifica I, 00133 Rome, Italy.}

\author{Emanuela Zaccarelli}
\email[Corresponding author: ]{emanuela.zaccarelli@cnr.it}
\affiliation{CNR-ISC and Department of Physics, Sapienza University of Rome,\\Piazzale A. Moro 2, 00185, Rome, Italy.}

\begin{abstract}
Poly-N-isopropylacrylamide (PNIPAM) phase diagram is explored in a wide range of temperature and pressure using extensive all-atom molecular dynamics simulations. By exploiting a simple model of an atactic linear polymer chain, we provide novel insights into PNIPAM coil-to-globule transition addressing the roles played by both temperature and pressure.
We find that a coil-to-globule transition exists up to large pressures, undergoing an intriguing reentrant behavior of the lower critical solution temperature with increasing pressure in agreement with experimental observations. Furthermore, we report the existence of a new kind of  globular state at high pressures, again confirming experimental results: this is characterized by a more structured hydration shell, that is closer to PNIPAM hydrophobic domains, with respect to the atmospheric pressure counterpart.  Our results clearly show that temperature and pressure induce PNIPAM coil-to-globule transition through different molecular mechanisms, opening the way for a systematic use of both thermodynamic parameters to tune the location of the transition and the properties of the associated swollen/collapsed states.
\end{abstract}

\maketitle

\section{Introduction}
The rational design of polymer materials that can fulfill targeted functions is at the frontier of research, notwithstanding the considerable progress made in macromolecular science in the last years~\cite{lutz2016precision}. Thermosensitivity of polymers in solution, whose thermodynamic foundations were laid by Flory in the middle of the last century~\cite{flory1953principles}, is still one the most investigated properties, being exploited in various nano and bio-technological applications~\cite{stuart2010emerging,fernandez2011microgel,agrawal2018functional,karg2019nanogels,di2020gellan}. Among synthetic polymers, Poly(N-isoproylacrylamide) (PNIPAM), is of particular interest because its lower critical solution temperature (LCST) in water at atmospheric pressure is about 305~K~\cite{halperin2015poly}, close to body temperature. The temperature-induced phase separation of PNIPAM aqueous solution was shown to occur with a coil-to-globule transition followed by the aggregation of polymer chains, while in PNIPAM-based chemically-crosslinked systems, the LCST behavior of linear chains gives rise to a volume phase transition (VPT)~\cite{fernandez2011microgel}. Hence, at atmospheric pressure, hydrogels and microgels of this polymer are found in a water-rich swollen state at ambient temperature, while, for temperatures above the LCST, water is partially expelled from the polymer network and the particles collapse~\cite{lopez2017does}. Moreover, at low temperatures, PNIPAM microgels were found to efficiently retain and confine water~\cite{zanatta2018}, with polymer-water interactions being the key ingredient driving PNIPAM solution behavior~\cite{Tavagnacco2018,niebuur2019water,tavagnacco2019water}.
Several studies using different experimental techniques have addressed the changes induced by temperature across the VPT, from both the structural~\cite{saunders2004structure,stieger2004small,conley2017jamming} and the dynamical~\cite{sierra2014structure,zanatta2020atomic} point of view, and more recently also the aid of numerical simulations turned out to be crucial in deeply understanding microgels phase behavior~\cite{rovigatti2019numerical}.

Pressure (P), like temperature (T), also induces a macroscopic phase separation of PNIPAM in aqueous solution~\cite{osaka2009quasi}, but its role has been rather less investigated so far and the molecular mechanisms occurring upon pressure changes have yet to be fully unveiled. More specifically, the $(P,T)$ PNIPAM phase diagram has been characterized in water~\cite{otake1993pressure,kunugi1997effects,osaka2009quasi} and in $D_2O$~\cite{niebuur2018formation}.
By exploiting the analogy of temperature-induced coil-to-globule transition of PNIPAM to the inverse process of protein cold denaturation, the increase of pressure is expected to induce an increase of the transition temperature, namely a stabilization of coil states~\cite{taniguchi1983studies,weingand1997combined,heinisch1995pressure}.
This is indeed observed also in PNIPAM for P$\lesssim$100~MPa~\cite{niebuur2018formation}. However, an opposite effect is detected at higher pressures where the temperature solubility range of PNIPAM chains is reduced. Such a reentrant phase behavior is also found in hydrogels~\cite{kato2005thermodynamic}: while up to pressures of about 150~MPa swollen states are stabilized up to temperatures about 10~K larger than the VPTT at atmospheric pressure, a further increase of pressure causes gradual shrinking. Moreover, a transition between a low pressure shrunken state and a high pressure shrunken state was proposed for hydrogels~\cite{kato2005thermodynamic}. This feature was confirmed by small angle X-ray scattering and Fourier transform infrared spectroscopy (FTIR) experiments for PNIPAM microgels where the collapsed states at high pressure turned out to be significantly more hydrated than those occurring at atmospheric pressure~\cite{grobelny2013,puhse2010influence}. The effect of size re-increasing by applying pressure is found to be independent on the polymer architecture~\cite{kunugi2005differences}. Since pressure seems to always favor the formation of more hydrated states~\cite{puhse2010influence}, the presence of a genuine coil-to-globule transition has been questioned at high pressures~\cite{meersman2005pressure}.

Altogether these findings show that pressure increases the water affinity of PNIPAM and plays an antagonistic effect with respect to temperature~\cite{grobelny2013}, with a balance that depends on the  specific $(P,T)$ conditions. It is thus important to perform an extensive study covering a wide range of T and P, in order to understand the concerted effects of these two state variables. Since experimentally pressure-induced aggregation may be a limitation for a quantitative investigation of the polymer structure and of its water environment~\cite{grobelny2013}, the use of molecular simulations is essential to  probe both intra- and inter-molecular interactions at the microscopic level. Surprisingly, up-to-date there have been no numerical studies on pressure effects on the PNIPAM coil-to-globule transition in water.
Here, we fill this gap by performing extensive all-atom molecular dynamics simulations of an atactic polymer chain at infinite dilution, exploring the PNIPAM phase diagram in a wide range of temperatures and pressures.
Our results reveal that temperature and pressure both induce a conformational transition of PNIPAM, but through different molecular mechanisms. This leads to a reentrant behavior of the coil-to-globule transition with increasing pressure and to the emergence of a novel `hydrated' globular state at high pressures, in qualitative agreement with experiments.

\section{Experimental}
PNIPAM phase diagram is investigated by performing all-atom MD simulations on an atactic linear polymer chain, made of 30 repeating units, at infinite dilution. PNIPAM and water are described with the OPLS-AA force field~\cite{Jorgensen1996} with the implementation by Siu et al.~\cite{Siu2012} and the Tip4p/ICE model~\cite{tip4pICE}, respectively, a simulation setup that was found to well reproduce PNIPAM coil-to-globule transition induced by temperature~\cite{Tavagnacco2018} and the effects due to the presence co-solvents~\cite{tavagnacco2020molecular}. Simulations are performed in a wide range of pressures between 0.1 and 500~MPa for temperatures between 283 K and 308 K, with a temperature step of 5 K. Trajectory data are collected for $\sim 0.3~\mu$s at each point in the P-T phase diagram. Further details on the simulation protocol are reported in sections I and II of the Supporting Information (SI) text.


\section{Results and discussion}
\begin{figure*}[th!]
\centering
\includegraphics[width=0.95\textwidth]{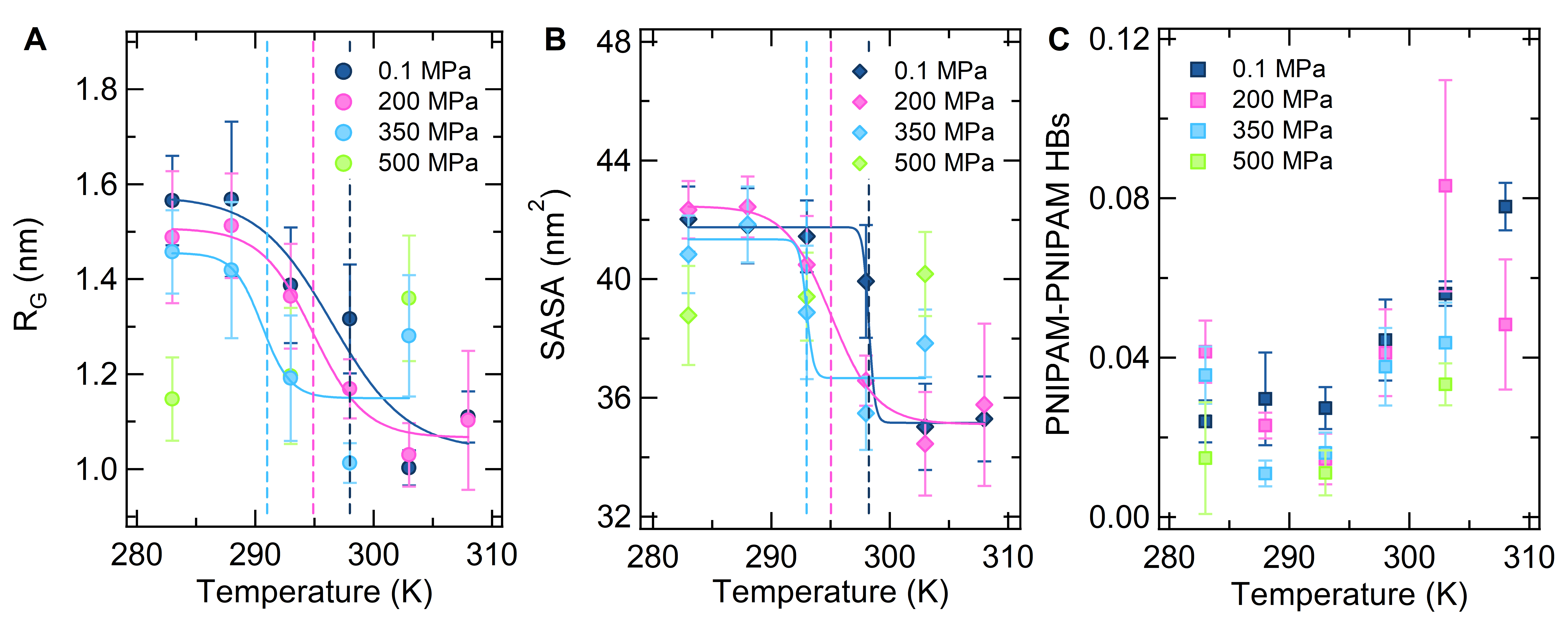}
  \caption{Temperature dependence of A) PNIPAM radius of gyration; B) solvent accessible surface area. Data represent time averaged values over the last 100 ns and standard deviation. Solid lines are the sigmoidal fit. Vertical dashed lines identify the LCST values. C) Temperature dependence of the number of PNIPAM-PNIPAM hydrogen bonds normalized to the number of repeating units. Data calculated at pressure values of 0.1, 200, 350 and 500~MPa are shown in blue, pink, light blue and green, respectively.}
  \label{fgr:pnipam}
\end{figure*}

We start by probing the high pressure regime (P $\geq$ 200~MPa) and examine the T-dependence of the average radius of gyration $R_G$ of the polymer chain. This observable signals the occurrence of the coil-to-globule transition, that is the infinite dilution manifestation of the phase separation taking place at the critical point.
Figure~\ref{fgr:pnipam}A reports $R_G$ as a function of temperature in the range between 283~K and 308~K for P= 0.1, 200, 350 and 500~MPa, while the corresponding time evolution for each studied state point is shown in Figure S1 of the SI text. For all pressure conditions except P=500~MPa, we  distinguish a temperature region where extended conformations are populated, attributable to coil states, and, at higher temperatures, the transition to the globule state.

Analysing the behavior of $R_G$, we notice three main features: (i) by increasing pressure the LCST, defined by the sigmoidal fit to the data in Figure~\ref{fgr:pnipam}A, decreases until no transition is observed for 500 MPa within the investigated T-range; (ii) for T < LCST the isothermal increase of pressure gradually reduces $R_G$ up to a collapse of the polymer chain at 500~MPa; (iii) for T > LCST the isothermal increase of pressure induces an increase of $R_G$. We further characterize the coil-to-globule transition of the polymer by calculating its average solvent accessible surface area SASA, displayed in Figure~\ref{fgr:pnipam}B, that fully reflects the behavior of $R_G$.  Notably, we observe in  Figure~\ref{fgr:pnipam}A that at 500~MPa the polymer chain adopts in the whole investigated T-range  a collapsed conformational state, which actually swells rather than deswells with increasing temperature. If we compare the high-P collapsed states with low-P ones at high T, we find larger values of $R_G$ and SASA at high pressure, suggesting a higher degree of hydration of the polymer chain. These results are consistent with experimental studies on PNIPAM hydrogels, microgels and linear chains~\cite{kato2005thermodynamic,puhse2010influence,lietor2011effect,grobelny2013}, validating our \emph{in silico} model, which can thus be further exploited to grasp important insights on the effects of pressure and temperature on PNIPAM solution behavior.

To this aim, we analyze the intramolecular interactions of the polymer chain. Each PNIPAM residue contains an amide group and thus hydrogen bonds can be formed between a CO and a NH group of different repeating units. In Figure~\ref{fgr:pnipam}C the average number of intrachain hydrogen bonds per polymer residue is reported as a function of temperature for different values of pressure. For T $\leq$ 293~K a small number of intramolecular hydrogen bonds is found, being weakly affected by changes in T and P. On the contrary, for T > 293~K temperature and pressure act differently: while the number of PNIPAM-PNIPAM hydrogen bonds is increased by temperature, it is reduced by pressure. The detected P-dependence is in agreement with evidence based on FTIR experiments~\cite{grobelny2013}. In addition, we directly compare the low and high pressure collapsed states and observe that the number of PNIPAM-PNIPAM hydrogen bonds at 500~MPa and 283~K is considerably lower than that at 0.1~MPa and 308~K, again indicating a higher exposition of amide groups to water at high pressure. However, we notice that the importance of these effects on the phase behavior is limited since the enthalpic contribution related to the formation of PNIPAM-PNIPAM hydrogen bonds, which favours globule conformational states, plays only a minor role in the coil-to-globule transition~\cite{pica2019does}.

\begin{figure*}[th!]
\centering
\includegraphics[width=0.95\textwidth]{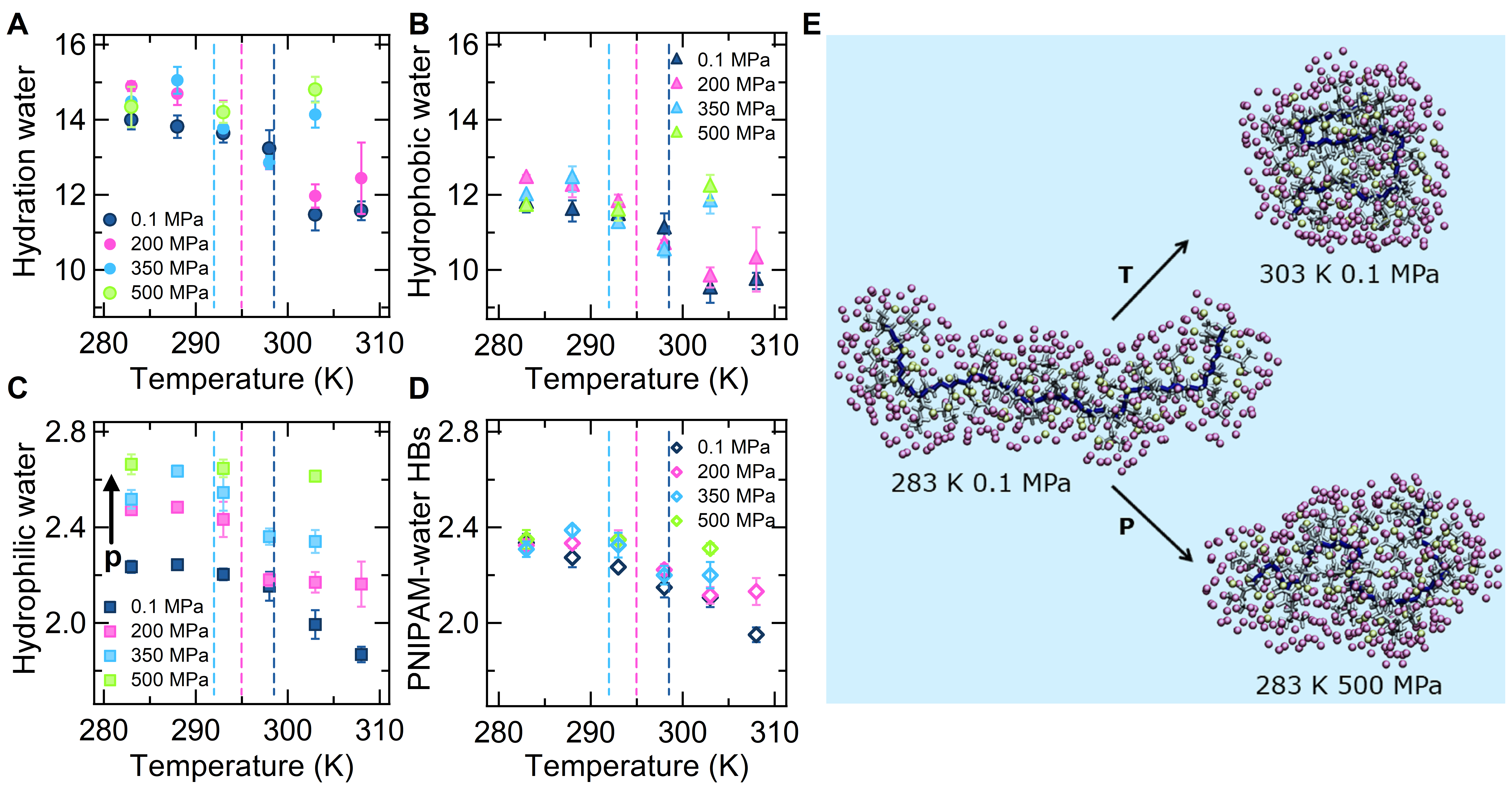}
  \caption{Temperature dependence of A) number of hydration water molecules; B) number of hydrophobic water molecules; C) number of hydrophilic water molecules and D) PNIPAM-water hydrogen bonds. Data calculated at pressure values of 0.1, 200, 350 and 500~MPa are shown in blue, pink, light blue and green, respectively. Numerical results are averaged over the last 100~ns of simulation and normalized to the number of repeating units. Vertical dashed lines mark the LCST values at 0.1~MPa (blue), 200~MPa (pink), and 350~MPa (light blue). E) Snapshots of a coil state at low $T$ and atmospheric pressure, a globule state at high $T$ and atmospheric pressure, a hydrated globule-like state at low $T$ and high pressure. PNIPAM atoms are shown in grey, with backbone carbons highlighted in blue. Oxygen atoms of hydration water are also displayed in pink and yellow when surrounding hydrophobic and hydrophilic groups, respectively.}
  \label{fgr:water}
\end{figure*}

We now turn our attention to the hydration properties of PNIPAM. In Figure~\ref{fgr:water}A we report the average number of hydration water molecules as a function of temperature and pressure, finding that the coil-to globule transition occurs with a drop of hydration water molecules at low pressures, although the polymer chain remains largely hydrated~\cite{pelton2010poly}. Importantly, such a drop is strongly reduced with increasing pressure. Actually, at 200 and 350~MPa and at a temperature sufficiently higher than the corresponding LCST, we detect a rehydration of the chain, not associated to a concomitant increase of SASA (Figure~\ref{fgr:pnipam}B). Furthermore, the number of hydration water molecules remains roughly constant with temperature at 500~MPa and the associated collapsed state at 283~K is characterized by a similar hydration degree to coil states occurring at the same temperature, but at lower pressures.

In order to evaluate how the local environment affects PNIPAM hydration, we compare the individual contributions of hydrophilic and hydrophobic hydration water, defined as water molecules mainly associated to the isopropyl moiety or to amide functional groups, respectively displayed in Figures~\ref{fgr:water}B and~\ref{fgr:water}C. A loss of both kinds of water is detected across the coil-to-globule transition. Qualitatively, hydrophobic water, which represents the major contribution, exhibits the same temperature/pressure behavior of the total hydration water. Instead, we find that temperature and pressure have an opposing effect on hydrophilic water: an isobaric temperature increase reduces its number while an isothermal increase of pressure raises it. We also notice that the collapsed states at high P have a larger number of hydrophilic water molecules with respect to the swollen state at atmospheric pressure. We further investigate the structuring of hydrophilic water molecules by calculating the average number of PNIPAM-water hydrogen bonds, reported in Figure~\ref{fgr:water}D. We observe that at ambient pressure, the number of PNIPAM-water hydrogen bonds is similar to the number of  hydrophilic  water molecules and thus all these molecules are directly interacting with the polymer. Differently, at high pressure, there is a significant amount of water molecules that surround amide groups, but do not form hydrogen bonds with PNIPAM, although the number of these bonds is large. Moreover, the associated clustering of hydration water molecules shows a reduction of average size upon isobaric temperature increase, whereas no net variation is detected upon isothermal pressure increase, even in the presence of globule states. However, the characteristic lifetime of PNIPAM-water hydrogen bond interactions decreases with pressure (see Figure S2 and Table S1 in the SI text).

Combining these results with the behavior of $R_G$ (Figure~\ref{fgr:pnipam}A), we observe that collapsed states at high and low P with comparable $R_G$ (e.g. 350 MPa at 298 K and 0.1 MPa at 303 K) clearly show different hydration properties. We thus provide direct evidence that pressure-induced globular conformations are not related to dehydration and that, despite an increased hydration, the coil-to-globule transition still occurs at high pressure, differently from the interpretation of Ref.~\cite{meersman2005pressure}. This is illustrated in the snapshots of Figure~\ref{fgr:water}E, which compares the coil state at low T and atmospheric pressure with the two different shrunken states: the standard globule and the so-called hydrated globule.

\begin{figure*}[th!]
\centering
\includegraphics[width=0.95\textwidth]{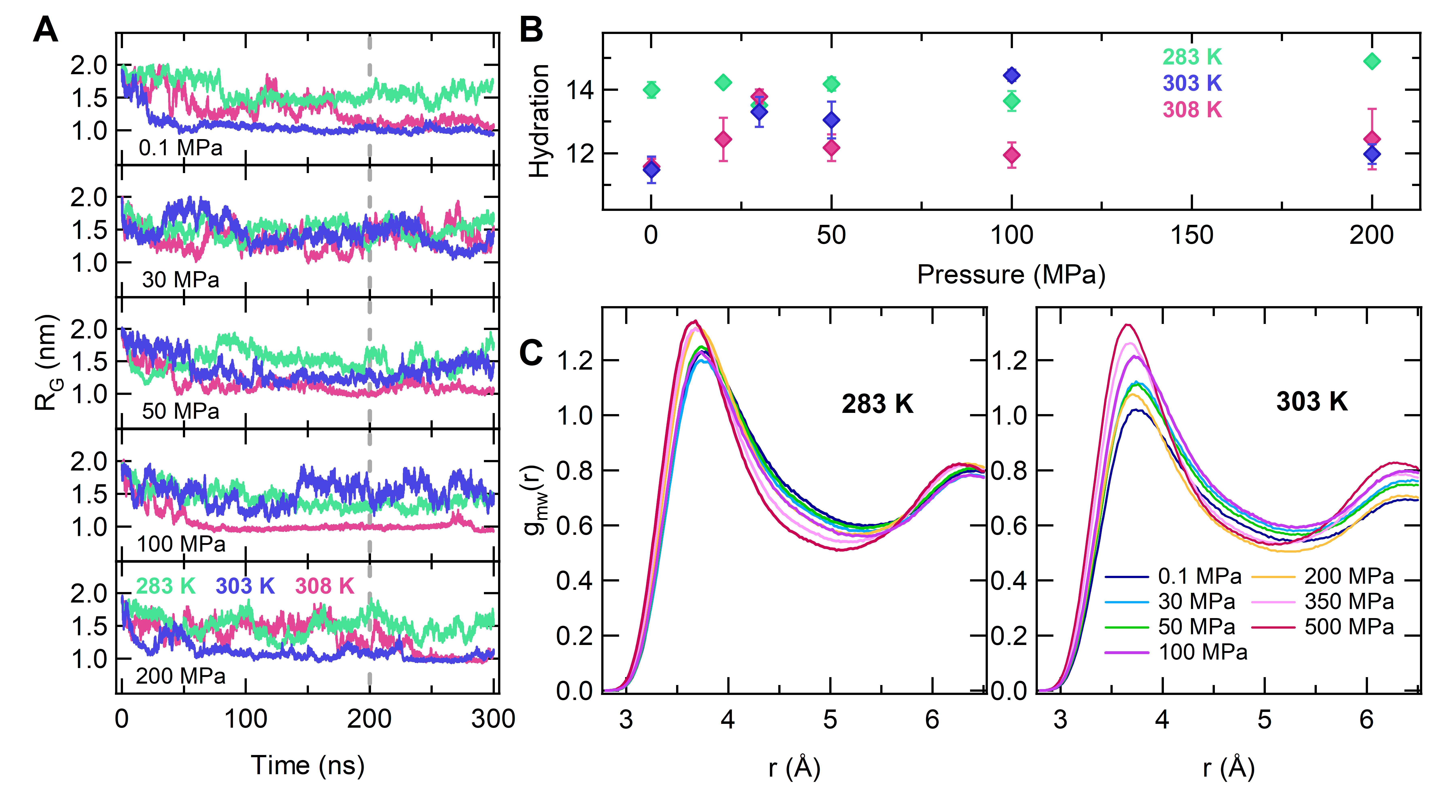}
  \caption{Reentrant behavior at low pressures: A) radius of gyration as a function of time for P=0.1, 30, 50, 100 and 200~MPa and B) number of hydration water molecules \emph{per} PNIPAM residue as a function of pressure. Data refer to T=283, 303 and 308~K; C) radial distribution functions for water oxygen atoms around PNIPAM methyl groups $g_{mw}$ for different pressures at T=283~K (left panel) and T=303~K (right panel). The analysis is performed over the last 100 ns trajectory interval, marked by the grey dashed vertical line in A.}
  \label{fgr:pressure}
\end{figure*}

\begin{figure}[th!]
\centering
\includegraphics[width=0.47\textwidth]{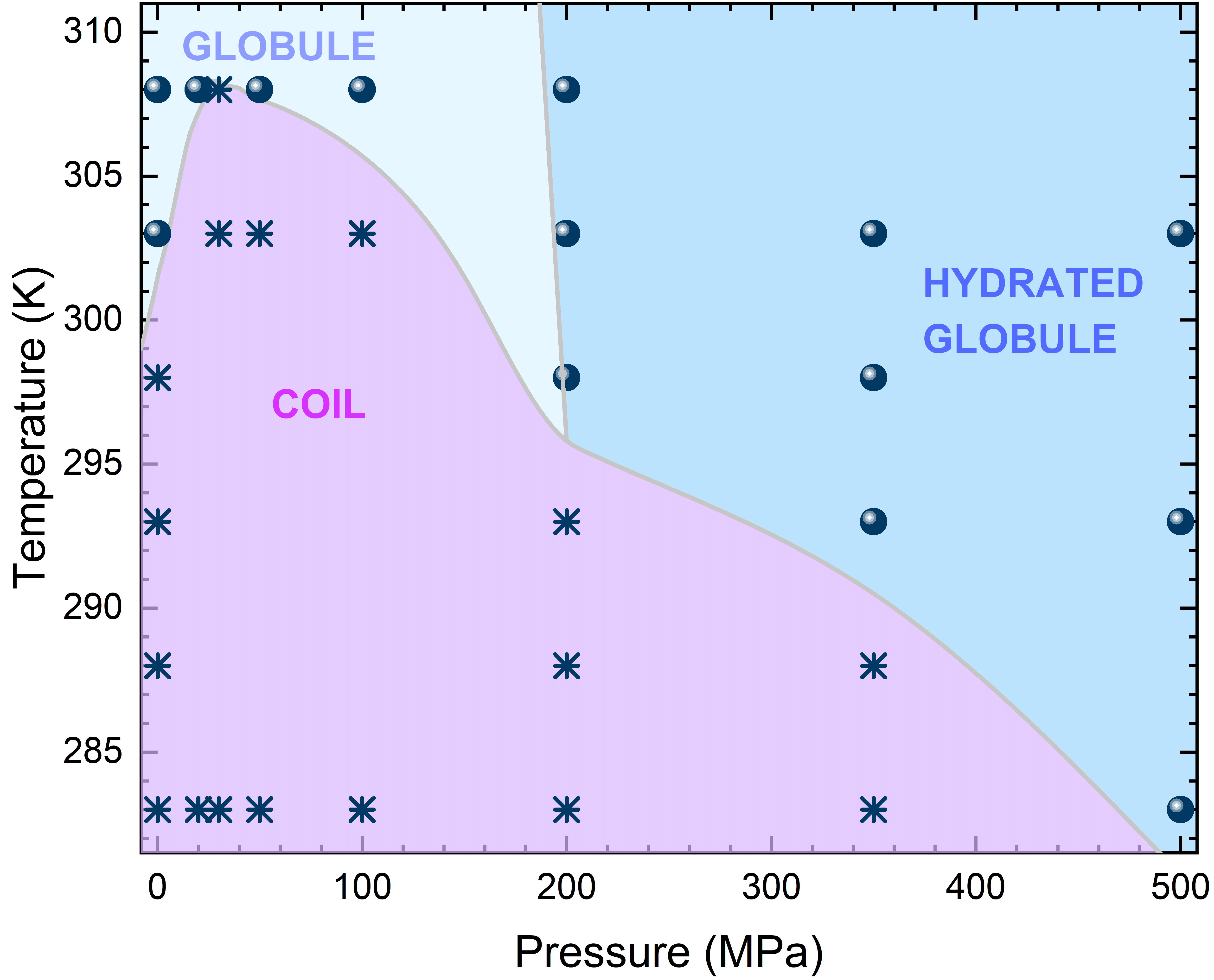}
  \caption{Summary of PNIPAM phase diagram. State points associated to coil and globule conformations are shown with asterisks and circles, respectively. Coloured areas are guide to the eye.}
  \label{fgr:summary}
\end{figure}

Having established PNIPAM behavior at high pressures, we now turn to examine the lower P domain (from atmospheric pressure up to 200 MPa) and the intriguing possibility to observe a reentrant deswelling transition. It is important to note that experimental evidence of this phenomenon is robust and independent of macromolecular architecture, though the range of pressures and temperatures where it is observed is very narrow. Hence, it is a big challenge for simulations to detect this feature, because of the need to scan a wide region of the phase diagram with enough resolution and the consequent very large computing costs.
Nonetheless, we report in Figure~\ref{fgr:pressure}A selected results for the chain $R_G$ at different pressures in the range 0.1-200~MPa for representative temperatures across the boundaries of the critical point. While we observe the transition to a collapsed state for 0.1~MPa and 200~MPa at T=303 and 308~K, we clearly detect the persistence of swollen states at 30, 50 and 100~MPa at 303~K. In addition, the chain also remains swollen at T=308~K and P=30~MPa, whereas it undergoes a coil-to-globule transition for 50 and 100~MPa at the same temperature. We can thus genuinely discriminate the occurrence of a non-monotonic variation of the LCST within the investigated time window. This finding is corroborated by the investigation of PNIPAM hydration properties for the same state points, that is reported in Figure~\ref{fgr:pressure}B. While the number of water hydration molecules is roughly constant in the studied P-interval for T=283~K, a non-monotonic behavior in pressure is observed at the two highest temperatures:
hydration has a maximum at P $\sim$ 100~MPa and at P $\sim$ 30~MPa for T=303~K and T=308~K, respectively. In both cases, hydration at first increases, then decreases and finally re-increases again for large pressures, as already shown in Figure~\ref{fgr:water}. The same behavior is found not only for total hydration water, but also for both hydrophobic and hydrophilic water molecules, as well as for PNIPAM-water hydrogen bonds (see Figure S3 in the SI text).
In order to obtain spatial information on hydration properties, we finally report the radial distribution functions between C atoms of side chain methyl groups and water oxygen $g_{mw}(r)$ in Figure~\ref{fgr:pressure}C. At T=283~K we find that the position of the main peak of $g_{mw}(r)$ is roughly constant up to P$\sim$100~MPa, moving toward smaller distances with a further increase of P. At the same time the peak height at first decreases, showing a minimum for P=30~MPa, and then continuously increases. Non-monotonic effects are even more evident for T=303~K, where the peak height increases up to 100 MPa and shows a marked decrease for 200~MPa, also revealing a change in the structuring of hydration water. A final increase of the peak height is detected in the high pressures regime. Notably, the features of the $g_{mw}(r)$ for globular conformations at high pressures indicate a stronger coordination of water to hydrophobic domains of PNIPAM, as experimentally detected~\cite{meersman2005pressure}.
These findings demonstrate the non-trivial interplay between temperature and pressure, which taken on their own would have antagonistic effects on the chain behavior: while a T-increase induces the chain to shrink reducing water contact, a P-increase promotes polymer hydration due to the more restricted volume of the tightly hydrated chain. However, when combined together, these two effects gives rise to unexpected behaviors. At pressures up to about 50 MPa, the gain of hydration water upon increasing pressure pushes the coil-to-globule transition at higher T. Then, at intermediate pressures  maximum hydration is reached and an overall reduction of chain size with a concomitant decrease of the LCST is detected. This implies that the two mechanisms driven by pressure and temperature are almost balanced for 100~MPa $\lesssim$ P $\lesssim$ 200~MPa, where the coil-to-globule transition occurs at the same temperature as for atmospheric pressure. For higher pressures, the decrease of the LCST persists, but with a change in the hydration mechanism, that is signalled by the decrease of the $g_{mw}(r)$ peak position, so that collapsed states for P$\gtrsim$ 200~MPa manifest as hydrated globules.  The resulting phase diagram, with all studied state points, is reported in Figure~\ref{fgr:summary}.

\section{Conclusions}
In conclusion, this work provides a comprehensive description of the molecular changes occurring in PNIPAM solution behavior when both temperature and pressure are varied. We have highlighted the complex pressure-dependence of the coil-to-globule transition, encompassing a non-monotonic dependence of the LCST with increasing pressure as well as the presence two different shrunken states at low and high pressures, respectively, in full qualitative agreement with experimental observations~\cite{niebuur2018formation,kato2005thermodynamic,grobelny2013}.

Finally, it is important to stress that the behavior of $R_G$ with increasing pressure along the T=303 K isotherm, corresponding to a pressure-induced rehydration of the globular conformation (Figs.~\ref{fgr:pnipam}A and~\ref{fgr:water}A) is analogous to the P-induced denaturation of globular proteins~\cite{mozhaev1996high}. As detected for pressure-denaturated proteins, at 303 K and high pressure PNIPAM retains a compact structure with water molecules penetrating the core~\cite{hummer1998pressure}. Although the molecular mechanism of the pressure induced denaturation  is still debated~\cite{kauzmann1987thermodynamics,royer2002revisiting,imamura2009effect,rouget2011size,roche2012cavities,mori2013pressure}, one of the main working hypothesis is the attenuation of the hydrophobic interactions between nonpolar side chains with increasing pressure~\cite{paliwal2004pressure,paschek2004reversible}, similarly to what detected in the present work for PNIPAM. Indeed the increased hydration of the hydrophobic regions of the polymer (Figure~\ref{fgr:pressure}C) witnesses the decrease of intrachain hydrophobic contacts. These findings thus also reinforce the role of PNIPAM as a useful model system to study hydration properties of proteins.


\begin{acknowledgements}
We acknowledge support from the European Research Council (ERC-CoG-2015, Grant No. 681597 MIMIC) and from MIUR (FARE Project No. R16XLE2X3L and SOFTART), and CINECA-ISCRA for HPC resources.
\end{acknowledgements}





%

\onecolumngrid
\renewcommand{\theequation}{S\arabic{equation}}
\renewcommand{\thefigure}{S\arabic{figure}}
\renewcommand{\thetable}{S\arabic{table}}
\renewcommand{\bibnumfmt}[1]{[S#1]}
\renewcommand{\citenumfont}[1]{S#1}
\setcounter{figure}{0}
\setcounter{section}{0}

\newpage

\section{Model and simulation procedure}
We investigate the pressure-temperature phase diagram of a PNIPAM linear chain in water in dilute regime. The degree of polymerization is chosen to be 30 on the basis of experimental results which showed that the solution behavior at atmospheric pressure of an oligomer made of 28 repeating units corresponds to that observed at higher degrees of polymerization~\cite{lutz2006,shan2009}. To reproduce the experimental conditions, we model the polymer chain with an atactic stereochemistry~\cite{chiessi2016,tavagnacco2018molecular}. Amide groups are represented with a trans geometry. PNIPAM and water are described with the OPLS-AA force field~\cite{jorgensen1996} with the modifications of Siu et al.~\cite{siu2012} and the Tip4p/ICE~\cite{tip4pICE} model, respectively.

First, the polymer chain with an energy optimized conformation~\cite{flory1966,chiessi2016} was centered in a cubic box of 8.5 nm side and oriented along a box diagonal to maximize the distance between periodic images. Then, 22849 water molecules were added and an energy minimization with tolerance of 1000 $kJ mol^{-1} nm^{-1}$ was carried out. The resulting system was used as initial configuration for the simulations at seven different pressure conditions, specifically 0.1~MPa, 30~MPa, 50~MPa, 100~MPa, 200~MPa, 350~MPa, and 500~MPa, in a range of temperature between 283~K and 308~K.
Trajectory acquisition and analysis were carried out with the GROMACS software package (version 5.1.4)~\cite{markidis2015solving,abraham2015gromacs}. The molecular viewer software package VMD was used for graphic visualization~\cite{humphrey1996vmd}.
MD simulations were carried out in the NPT ensemble for 300 ns for each point in the P-T phase diagram. Trajectories were acquired with the leapfrog integration algorithm~\cite{hockney1970potential} with a time step of 2 fs. Cubic periodic boundary conditions and minimum image convention were applied. The length of bonds involving H atoms was constrained by the LINCS procedure~\cite{hess1997lincs}. The velocity rescaling thermostat coupling algorithm, with a time constant of 0.1 ps was used to control temperature~\cite{bussi2007canonical}. Pressure of was maintained by the Berendsen barostat~\cite{berendsen1984} using a time constant of 0.5 ps with a standard error on pressure values lower than 10\%. The cutoff of nonbonded interactions was set to 1 nm and electrostatic interactions were calculated by the smooth particle-mesh Ewald method~\cite{essmann1995smooth}. Typically, the final 100 ns of trajectory were considered for analysis, sampling 1 frame every 5 ps.

\section{Trajectory analysis}
The occurrence of the coil-to globule transition was monitored by calculating the radius of gyration ($R_G$) and the solvent accessible surface area (SASA). $R_G$ was calculated through the equation:

\begin{eqnarray}
R_g&=&\sqrt{\frac{\sum_{i}\|r_i\|^2m_i}{\sum_{i}m_i}}
\label{Eq:rg}
\end{eqnarray}

\noindent where $m_i$ is the mass of the $i^{th}$ atom and $r_i$ the position of the $i^{th}$ atom with respect to the center of mass of the polymer chain. The SASA is defined as the surface of closest approach of solvent molecules to a solute molecule, where both solute and solvent are described as hard spheres. It is calculated as the van der Waals envelope of the solute molecule extended by the radius of the solvent sphere about each solute atom centre~\cite{richmond1984solvent}. We used a spherical probe with radius of 0.14 nm and the values of Van der Waals radii of the work of Bondi~\cite{bondi1964van,eisenhaber1995double}. The distributions of SASA values were calculated with a bin of 0.1 nm$^2$. Averages were performed over the last 100 ns of trajectory.

Hydrogen bonds were evaluated adopting the geometric criteria of an acceptor-donor distance ($A \cdot \cdot \cdot D$) lower than 0.35 nm and an angle $\theta$ ($A \cdot \cdot \cdot D-H$) lower than $30^\circ$, irrespective of the $AD$ pair. The dynamical behavior of hydrogen bonding interactions was characterized by calculating the normalized intermittent time autocorrelation function which is irrespective of intervening interruptions. The corresponding lifetime was defined as the time at which the autocorrelation function is decayed of the 50\% of its amplitude to reduce statistical noise.

Hydration properties were investigated as a function of pressure and temperature. The ensemble of water molecules in the first solvation shell was sampled by selecting molecules having the water oxygen atom at a distance from PNIPAM nitrogen/oxygen atoms lower than 0.35 nm (hydrophilic water molecules) or a distance from methyl carbon atoms of PNIPAM lower than 0.55 nm (hydrophobic water molecules). Hydrogen bonding between water molecules of the first solvation shell was also analysed. The clustering of water molecules belonging to the first solvation shell was analysed by calculating the population of clusters formed by hydrogen bonded molecules.

\section{Structural characterization of PNIPAM coil-to-globule transition}
\begin{figure}[th!]
    \includegraphics[width=0.95\textwidth]{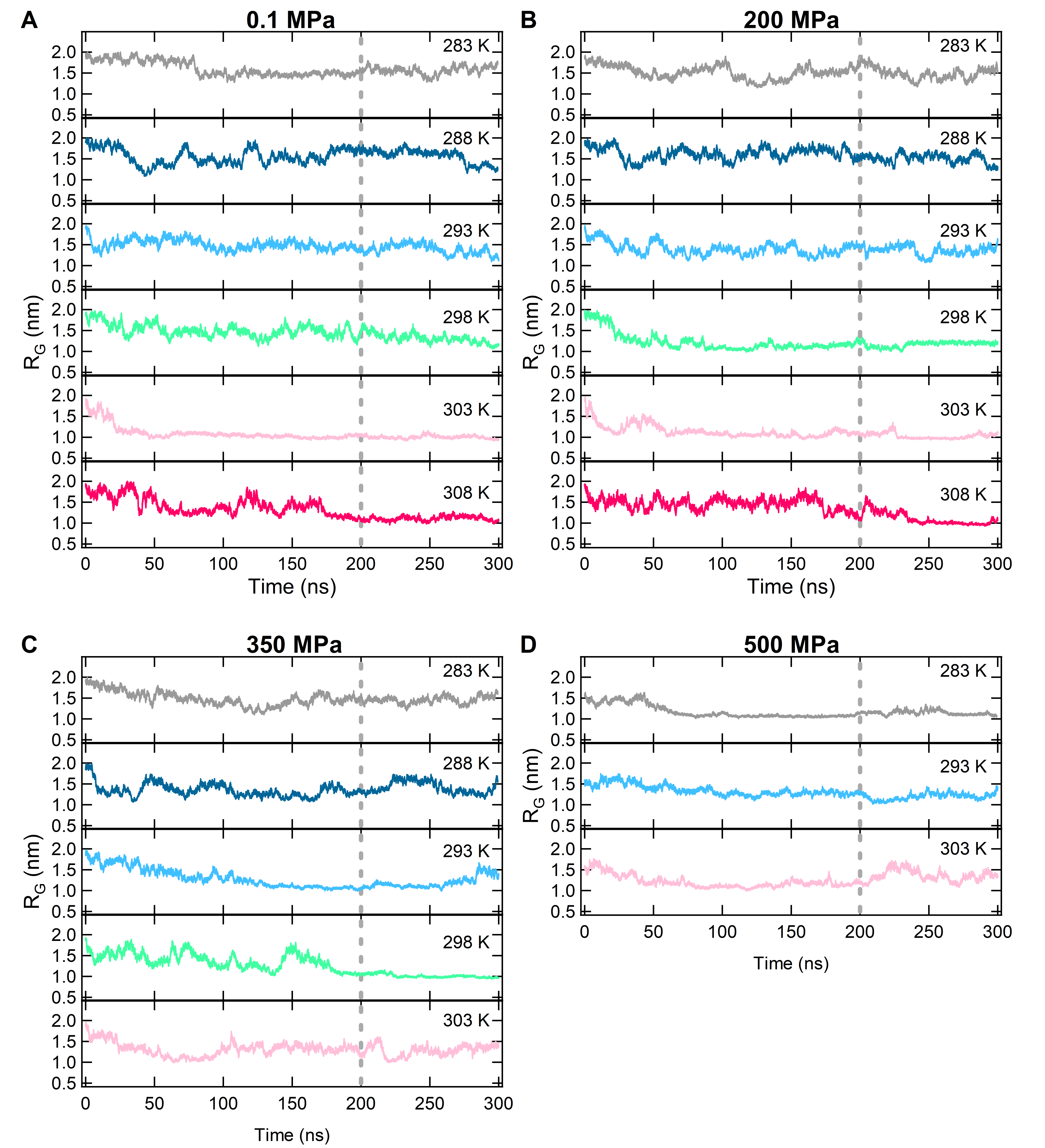}
    \caption{Time evolution of PNIPAM radius of gyration as a function of temperature at the investigated pressure values of 0.1 MPa (\textbf{A}), 200 MPa (\textbf{B}), 350 MPa (\textbf{C}), and 500 MPa (\textbf{D}). Simulation results obtained at 283 K, 288 K, 293 K, 298 K, 303 K, and 308 K are displayed in gray, blue, light blue, green, pink, and red, respectively. The last 100 ns of trajectory data, marked by dashed vertical lines, were used for data analysis.}
    \label{fig:RG}
\end{figure}

To identify the lower critical solution temperature (LCST), we calculated the radius of gyration of the polymer chain in the last 100 ns of equilibrated run. Figure~\ref{fig:RG} shows the time evolution of the radius of gyration of the chain in the whole trajectory for the simulations at atmospheric pressure and in the high pressure regime.

\newpage

\section{Water molecules clustering}
\begin{figure}[th!]
\centering
\includegraphics[width=12 cm]{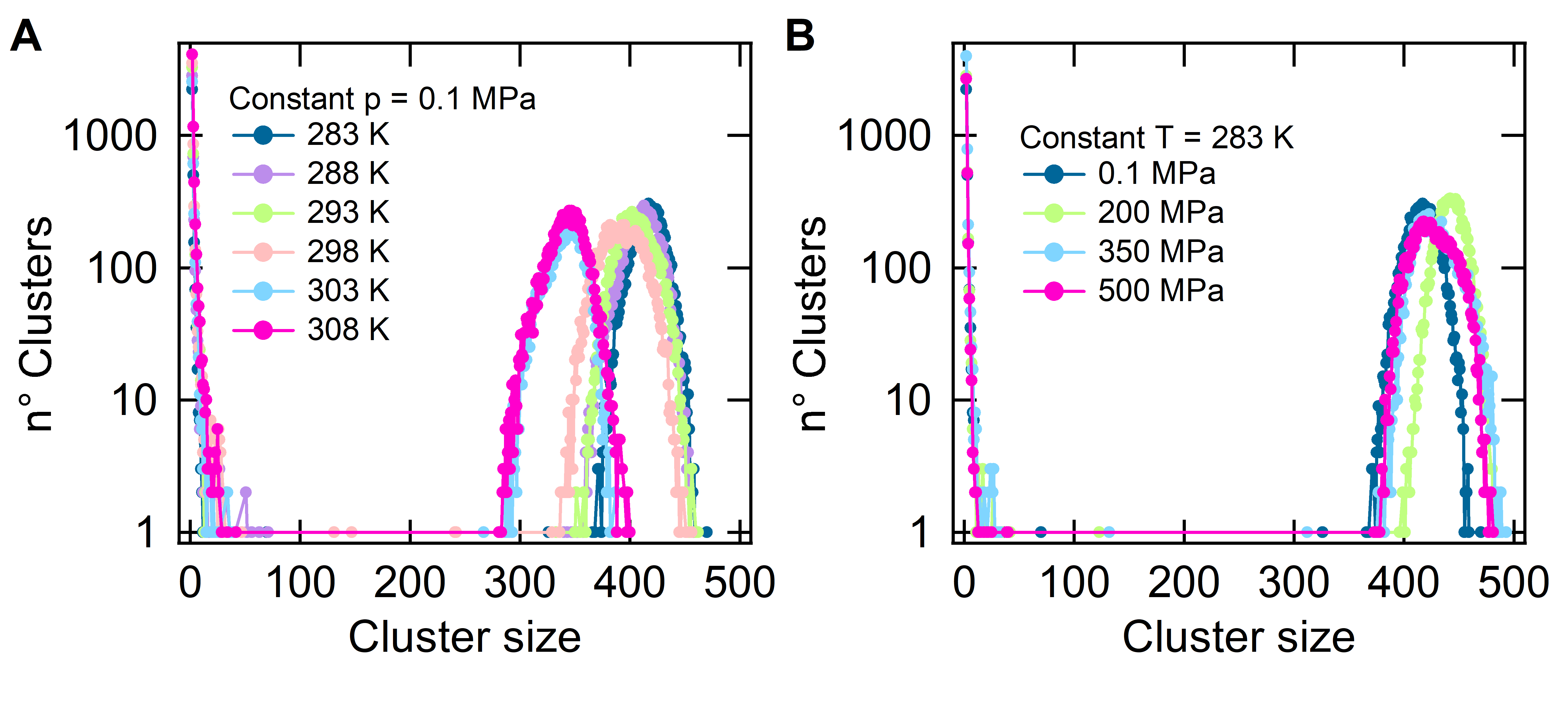}
  \caption{Distribution of cluster sizes of water molecules in the first solvation shell of the polymer chain: A) isobaric behavior at P= 0.1~MPa and B) isothermal behavior at T=283~K. Data obtained at 0.1~MPa and 283~K, 288~K, 293~K, 298~K, 303~K, and 308~K are shown in blue, violet, green, orange, light blue and pink, respectively. Data calculated at 283~K and 0.1~MPa, 200~MPa, 350~MPa, and 500~MPa are displayed in blue, green, light blue, and pink, respectively.}
  \label{fgr:cluster}
\end{figure}

To better understand the differences observed in PNIPAM hydration features, we have characterized the clustering of water molecules within the first solvation shell, defined as the number of water molecules interconnected through hydrogen bonds. Figure~\ref{fgr:cluster}A and Figure~\ref{fgr:cluster}B compare the changes occurring in the distribution of cluster sizes of water molecules when PNIPAM coil-to-globule transition is induced by temperature or pressure. At 283~K and 0.1~MPa we observe the presence of a large number of small clusters and a distribution of values centered at about 420 water molecules. By heating the system at 0.1~MPa a clear reduction of the average size of the distribution is observed when PNIPAM undergoes the coil-to globule transition. Differently, when pressure is increased at 283~K no such net variation is detected, which denotes a stability of the structure of the hydration shell, although the coil-to-globule transition occurs above 350 MPa. Therefore Figure~\ref{fgr:cluster} shows that P-induced globule states are characterized by a more structured hydration shell as compared to T-induced globule states.

\section{PNIPAM-water hydrogen bonding interactions}
We investigated the changes in PNIPAM hydration properties through the characterization of the characteristic lifetime of PNIPAM-water hydrogen bonding interactions, as reported in Table~\ref{tbl:lifetime}. We find a reduction of the lifetime by applying pressure, while the transition to globular conformation slightly increases it. These results are coherent with the increased mobility of PNIPAM hydration water moving from 0.1 to 130 MPa, as well as the reduction of mobility after the phase separation, as experimentally detected~\cite{niebuur2019water}.

\begin{table}[th!]
  \centering
  \caption{Lifetime of PNIPAM-water hydrogen bonds.}
  \label{tbl:lifetime}
  \begin{threeparttable}
  \setlength{\tabcolsep}{16pt}
  \begin{tabular}{ccccc}
    \hline
    T (K) & 0.1~MPa & 200~MPa & 350~MPa  & 500~MPa  \\
    \hline
    283  & 126 ($\pm$8) &  94 ($\pm$10) &  87 ($\pm$10) & 102 ($\pm$11)$^*$ \\
    288  &  74 ($\pm$6) &  67 ($\pm$5)  &  65 ($\pm$8)  &              \\
    293  &  65 ($\pm$3) &  66 ($\pm$11) &  66 ($\pm$4)$^*$  &  84 ($\pm$13)$^*$ \\
    298  &  50 ($\pm$4) &  69 ($\pm$6)$^*$  &  64 ($\pm$11)$^*$ &              \\
    303  &  63 ($\pm$5)$^*$ &  50 ($\pm$5)$^*$  &  48 ($\pm$6)$^*$  &  44 ($\pm$5)$^*$  \\
    308  &  41 ($\pm$4)$^*$ &  41 ($\pm$9)$^*$  &               &              \\
    \hline
    \end{tabular}
        \begin{tablenotes}
        \footnotesize
        \item \emph{Lifetimes are reported in ps. Errors are estimated by applying the blocking method. Lifetimes associated to globular conformations are marked with an asterisk.}
        \end{tablenotes}
    \end{threeparttable}
\end{table}

\section{PNIPAM hydration properties at pressures below 200 MPa}
\begin{figure}[th!]
\centering
\includegraphics[width=8 cm]{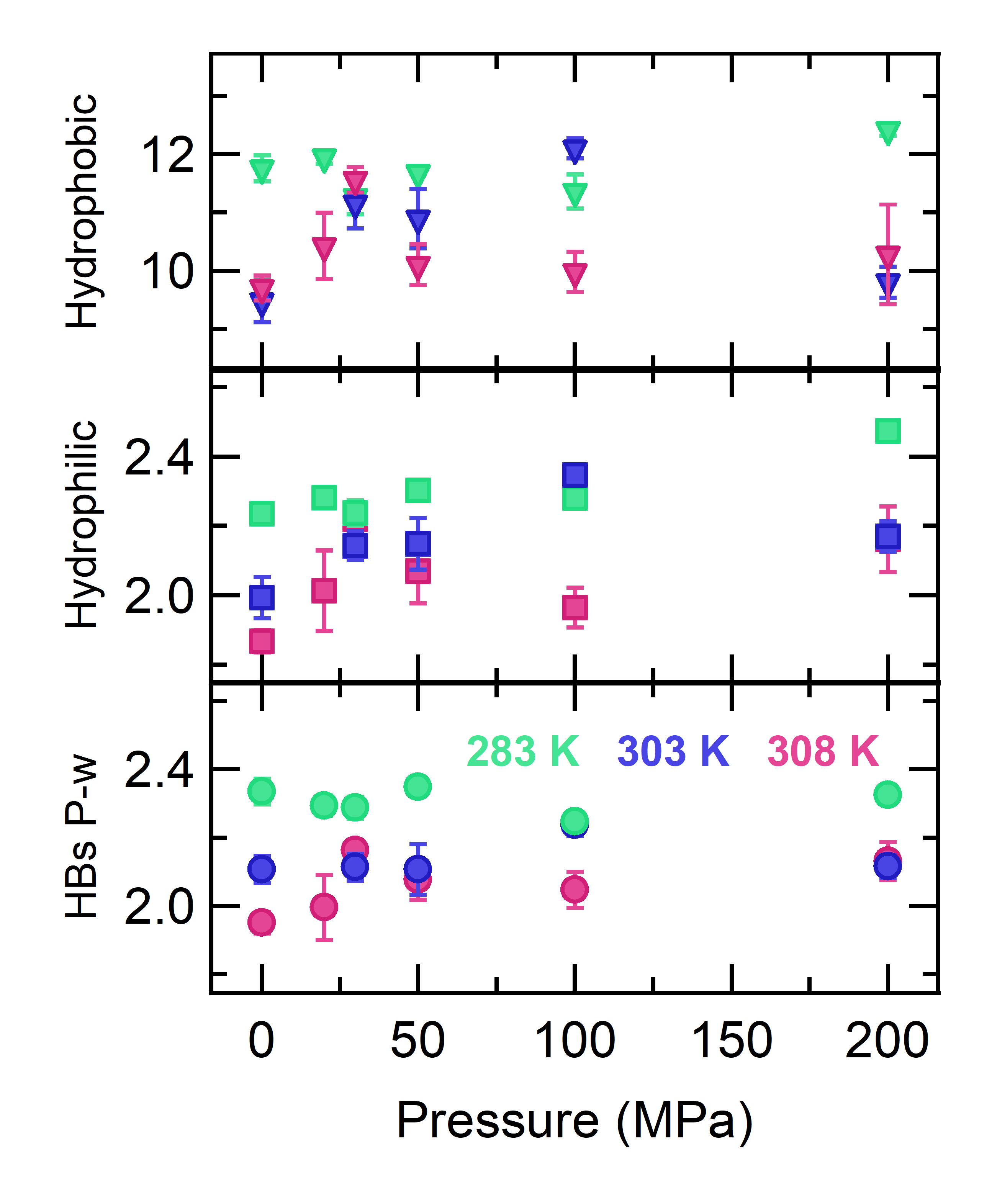}
  \caption{Average number of hydrophobic and hydrophilic water molecules \emph{per} PNIPAM residue and PNIPAM-water hydrogen bonds (HBs P-w) \emph{per} PNIPAM residue as a function of pressure. Data are shown at the boundaries of the critical point for T=283 (green), 303 (blue) and 308~K (red).}
  \label{fgr:lowp}
\end{figure}

We monitored the changes occurring in PNIPAM hydration properties in between 0.1 and 200 MPa by evaluating the individual contribution of hydrophobic and hydrophilic water molecules in the first solvation shell of the polymer chain, along with the number of PNIPAM-water hydrogen bonds, as reported in Figure~\ref{fgr:lowp}. At T=303~K and T=308~K all these properties exhibit a non-monotonic behavior in pressure, characterized by a first increase and then a decrease at intermediate P. At 308 K a final re-increase at larger pressure is also visible.

\end{document}